\documentclass[conference]{IEEEtran}
\IEEEoverridecommandlockouts
\usepackage{cite}
\usepackage{amsmath,amssymb,amsfonts}
\usepackage{algorithmic}
\usepackage{graphicx}
\usepackage{subfig}
\usepackage{textcomp}
\usepackage{xcolor}
\usepackage{listings}
\usepackage{hyperref}
\def\BibTeX{{\rm B\kern-.05em{\sc i\kern-.025em b}\kern-.08em
    T\kern-.1667em\lower.7ex\hbox{E}\kern-.125emX}}
\begin{document}

\title{Towards Transactional Causal Consistent Microservices Business Logic}

\author{\IEEEauthorblockN{Pedro Pereira and António Rito Silva}
\IEEEauthorblockA{INESC-ID, Instituto Superior Técnico, University of Lisbon -- Lisbon, Portugal \\
\{pedro.l.pereira, rito.silva\}@tecnico.ulisboa.pt}
}

\maketitle
\thispagestyle{plain}
\pagestyle{plain}
\begin{abstract}
Microservices architecture has been widely adopted to develop software systems, but some of its trade-offs are often ignored. In particular, the introduction of eventual consistency has a huge impact on the complexity of the application business logic design. Recent proposals to use transactional causal consistency in serverless computing seems promising, because it reduces the number of possible concurrent execution anomalies that can occur due to the lack of isolation. We propose an extension of the aggregate concept, the basic building block of microservices design, that is transactional causal consistent compliant. A simulator for the enriched aggregates is developed to allow the experimentation of this approach with a business logic rich system. The experiment results shown a reduction of the implementation complexity. The simulator is a publicly available reusable artifact that can be used in other experiments.
\end{abstract}


\section{Introduction}
\label{sec:introduction}

Microservices have become increasingly adopted~\cite{Hanlon06} as the architecture of business systems, because it promotes the split of the domain model into consistent pieces that are owned by small agile development teams, and facilitates scalability~\cite{thones15,fowler_microservices}. 

These systems are implemented using the Saga pattern~\cite{richardson19} to handle the concurrency anomalies, like lost update and dirty reads, which result on extra complexity in the implementation of the system business logic~\cite{Santos20}. It has been identified a trade-off between the business logic complexity and use of microservices~\cite{Haywood17}, which is also confirmed by the type of systems where the use of microservices has been successfully reported, where there is the need for high scalability, but the domain business logic is less complex, e.g. Netflix.

Recent research has proposed the use of transactional causal consistency to support serverless computing~\cite{Wu20,lykhenko21}, which reduce the number of anomalies by providing a causal snapshot to support the functionalities distributed execution. However, as far as we know, there is no experimentation of this approach with systems that have complex business logic. On the other hand, the two implementations provide a low level API which is not friendly for the software developer.

Therefore, we intend to answer the following research questions:

\begin{enumerate}
    \item Does the use of transactional causal consistency simplify the microservices implementation of business logic rich systems?
    \item Does the use of a transactional causal consistency simulator ease the experimentation with large domain models?
\end{enumerate}

To answer the research questions, we extend the domain-driven concept of aggregate~\cite{evans03}, which is the basic building block of microservices systems~\cite{richardson19}, with causal consistency semantics. Afterwards, it is designed, and implemented, a simulator for aggregates enriched with transactional causal consistency. Then, a large software system, with rich business logic, is implemented. Finally, the results are evaluated.

As result of this work, a set of constructs are proposed to enrich aggregates, a transactional causal consistency simulator is made available, and guidelines on how to implement business logic with transactional causal consistency are defined.

After the introduction in this section, related work is presented in Section~\ref{sec:relatedWork}. The constructs for aggregate definition are presented in Section~\ref{sec:aggregates}, which are extended in Section~\ref{sec:semantics}, that defines the semantics of transactional causal consistency for microservices systems built with aggregates. Section~\ref{sec:simulator} describes the simulator design and implementation. Section~\ref{sec:caseStudy} describes the case study and Section~\ref{sec:evaluation} analysis of the implementation of the case study using the simulator. Finally, Section~\ref{sec:conclusions} draws the conclusions.

\section{Related Work}
\label{sec:relatedWork}

Microservices architecture has to comply with what is stated by the CAP theorem~\cite{Fox99}, in particular, consistency has to be traded-off for availability. Therefore, eventual consistency~\cite{bailis13} has been adopted in the implementation of microservices, using sagas~\cite{Garcia-Molina87,richardson19}. However, writing application business logic in the context of the eventual consistency requires an extra effort~\cite{Santos20} to deal with anomalies like lost updates and dirty reads. This is due to the intermediate states created by the functionality execution in each one of the microservices. Due to this lack of isolation, the business logic is intertwined with the handling of incomplete states associated with the concurrent execution with other functionalities. This complexity depends on the number of these intermediate states, which depend on the number of functionalities~\cite{Santos20}.

Therefore, implementing a system using microservices is not trivial, depends on the complexity of the business logic~\cite{Haywood17} and sometimes is too complex and systems are even migrated back to a monolith architecture~\cite{Mendonca21}.

Transactional causal consistency (TCC)~\cite{akkoorath16} has been proposed as a transactional model that handles some of the problems of eventual consistency, by providing to the executing functionality a consistent causal snapshot. The entities in the causal snapshot respect the \textit{happens-before} relation and the writes performed when the functionality commits are atomically visible, regardless of node in which read or write operation occurs. This transactional model handle dirty reads, because the reads are consistent, but continue to allow lost updates, which occur when two concurrent functionalities write the same entity, the last to commit overwrites the first one. On the other hand, transactional causal consistency can be implemented using non-blocking algorithms, overcoming the limitations stated in the CAP theorem.

As far as we know, there are two implementations of TCC for serverless computing~\cite{Wu20,lykhenko21}, but these implementations use a key-value store and/or offer a low level API that do not facilitate to experiment the support of complex business logic with TCC. Additionally, they only use toy cases to experiment the application of transactional causal consistency.

On the other hand, the design of microservices is based on the domain-driven design concept of aggregate~\cite{evans03,richardson19}, which denotes the transactional unit of consistency in microservices. This concept is not considered by the existing implementations of TCC, which focus on the management of replicas, instead of the different perspectives of a same model in different bounded contexts~\cite{evans03}. However, there are some synergies between both concepts that are worth exploring, but, as far as we know, are not addressed by the literature.

There are some research on the extension of aggregates to ensure in data-intensive distributed systems, such as microservices, consistency between replicas of entities~\cite{braun21}. However, these extensions focus on microservices systems running using eventual consistency. Other relevant work that deals with the replica consistency is the conflict-free replicated data types research~\cite{Preguica2009,yu20}, which use operations for replica consistency. However, this type of approach requires a completely different perspective on how software developers write code, and on the middleware infrastructure, which has to support the concept of operation.

Therefore, we intend to leverage on existing work by enriching the concept of aggregate to support TCC semantics. Additionally, develop a TCC simulator that eases the experiment with the implementation of complex microservices systems, applying TCC semantics to the functionalities execution.

\section{Aggregates}
\label{sec:aggregates}

Aggregates are considered the basic building blocks of microservices applications~\cite{richardson19}. The concept was imported from domain-driven design~\cite{evans03} and defines a unit of consistency between the aggregate entities, which is defined as the aggregate invariants. An aggregate has a root entity that controls its accesses to guarantee atomicity and aggregate internal entities are not visible from outside. In the context of microservices, aggregate accesses occur in the context of ACID transactions. 

When splitting a domain model into aggregates it is necessary to consider the consistency between aggregates. In domain-driven design the consistency between aggregates can be relaxed. To distinguish these two types of consistency, we define intra-invariants and inter-invariants. The former define a rule on the aggregate entities while the latter between different aggregates. We also consider a upstream-downstream relation between aggregates, which is similar to the same relation between bounded contexts in domain-driven design. An inter-variant is defined in the downstream aggregate, because it is aware of the upstream aggregate model, while the opposite does not occur.

\begin{minipage}{\linewidth}
\begin{lstlisting}[language=Java, basicstyle=\tiny, caption=Tournament Aggregate, label={lst:tournamentAggregate}]

Aggregate Tournament
    Integer id
    LocalDateTime startTime, endTime
    List<Participant> participants
    ...
    Participant is s: CourseExecution.students
        id from s.id
        name from s.name
    ...
    Intra-Invariants
        START_BEFORE_END
	this.startTime < this.endTime
    ...
    Inter-Invariants
        PARTICIPANT_EXISTS
	forall p : this.tournamentParticipants | p.state != INACTIVE => 
            exists s = CourseExecution.students(p.id)
            p.name == s.name
        ...
\end{lstlisting}
\end{minipage}

Listing~\ref{lst:tournamentAggregate} contains the representation of an excerpt of a tournament aggregate. The root entity is the tournament, which has a unique identifier. Attributes \textit{startTime} and \textit{endTime} represent the period the tournament is open. Internal entity \textit{participants} represents students participating in the tournament. The \textit{Participant} entity is associated with a student enrolled in a course, which is an element of the upstream aggregate (\textit{CourseExecution}), and contains the student identification and name. Therefore, a consistency issue may occur between these two aggregates, which is declared by the \textit{PARTICIPANT\_EXISTS} inter-invariant. Finally, \textit{START\_BEFORE\_END} declares a intra-aggregate invariant, which should be preserved whenever the aggregate is changed.

\begin{minipage}{\linewidth}
\begin{lstlisting}[language=Java, basicstyle=\tiny, caption=Tournament Functionalities, label={lst:tournamentFunctionalities}]

Functionality updateDates
    update this.startTime
    update this.endTime

Functionality anonymizeParticipant(p)
    update this.participants(p.id).name
\end{lstlisting}
\end{minipage} 

Aggregates have functionalities. In listing \ref{lst:tournamentFunctionalities} declares two tournament functionalities: \textit{updateDates} that updates the tounament open period; and \textit{anonymizeParticipant} that anonymizes a participant. The latter is related with the \textit{PARTICIPANT\_EXISTS} inter-invariant, while the former with the \textit{START\_BEFORE\_END} intra-invariant.

\section{Semantics}
\label{sec:semantics}

\subsection{Functionalities}

To define the aggregates semantics in a microservices architecture, when the functionalities are executed using transactional causal consistency, it is necessary to classify the types of functionalities that exist. A functionality is associated with an aggregate, referred as the functionality aggregate or the main aggregate, where it preforms reads and/or writes. A functionality can perform reads and writes on other aggregates, besides the main aggregate, they are referred as the functionality secondary aggregates. Finally, due to architectural upstream-downstream relationship between aggregates, a functionality can directly perform reads and writes on the upstream aggregates of its main aggregate, but cannot do any type of access to its downstream aggregates. Therefore, the secondary aggregates of a functionality have to be upstream aggregates of its main aggregate. Nevertheless, a functionality can publish events that may be subscribed by downstream aggregates, which corresponds to a kind of indirect write, if the downstream aggregate uses the event to change its state, but the initiative to perform the change is on the downstream aggregate, which is aware of the semantics of its upstream aggregates. Note that, a functionality cannot perform a read on a downstream aggregate, because publishing events is asynchronous, and, from an architectural perspective, the upstream aggregate is unaware of downstream aggregates. 

Therefore, considering these concepts, the functionalities can be classified in the following types:
\begin{itemize}
    \item \textit{Query}: this type of functionality only contains read operations to aggregates, and is distinguished by the number of aggregates it reads:
        \begin{itemize}
            \item \textit{Single Aggregate} - all reads belong to the same aggregate, the main aggregate;
            \item \textit{Multiple Aggregates} - the reads are done in several aggregates, the main aggregate and one or more secondary aggregates.
        \end{itemize}
    \item \textit{Simple Functionality}: characterizes the functionalities that write in a single aggregate, the main aggregate, though they may read different aggregates, and so, it is also distinguishes by the number of aggregates it reads: single and multiple.
    The write on the main aggregate may trigger events to be subscribed by downstream aggregates;
    \item \textit{Complex Functionality}: writes in multiple aggregates, the main aggregate and one or more secondary upstream aggregates. Like in the previous type, it can read downstream aggregates: single or multiple.
    Additionally, the writes on the aggregates may trigger events to be subscribed by downstream aggregates.
\end{itemize}

\subsection{Transactional Causal Consistency}

To define the semantics of the execution of functionalities in the context of transactional causal consistency, we introduce the concept of version number. Each aggregate has several versions, where $A$ is the set of all aggregate version, and each version has a unique number, denoted by $a.version$, where $a \in A$. The version numbers form a total order, i.e. it is possible to compare any two version numbers, $\forall_{a_i,a_j \in A} a_i.version \leq a_j.version \lor a_j.version < a_i.version$. Given an aggregate version, $a \in A$, $a.aggregate$ denotes its aggregate, and $a.versions$ denotes all the aggregates versions of aggregate $a.aggregate$.
Additionally, there is a version number for each functionality $f$, denoted by $f.version$, that is assigned when the functionality starts with the number of the last successfully finished functionality incremented by one. This number may be subsequently changed, as it will be described, and assigned as the version number of all the aggregates written by the functionality. $F$ is the set of all executed functionalities, and $F.success \subseteq F$ is the subset of functionality executions that finished successfully. 

A causal snapshot of an executing functionality is a set of aggregates which are causality consistent given the functionality version number. Therefore, a functionality $f$ causal snapshot, denoted by $f.snapshot$, is a set of aggregate versions, $f.snapshot \subseteq A$, such that there is a single version for each aggregate, $\nexists_{a_i,a_j \in f.snapshot} a_i.aggregate = a_j.aggregate$, and the following condition holds:
\begin{equation*}
\begin{split}
    & \forall_{a_i \in f.snapshot}: a_i.version < f.version \text{ } \land \\
    & \forall_{a_j \in a_i.versions}: a_j.version < a_i.version
\end{split}
\end{equation*}

The first condition guarantees that the version was not created by a functionality that finished after $f$ started, and the second condition guarantees that it is the most recent version, considering the first condition. As an example, consider an aggregate version $a$ which as version number 5, it was written by functionality with version number 5. Also consider two functionalities, $f_i$ and $f_j$, that start concurrently, when the last successfully finished functionality have 7 as its version number. So, the version number of $f_i$ and $f_j$ is 8. Suppose the $f_i$ finishes first and writes a version of $a$, which will have version number 8. If $f_j$ reads $a$ to its snapshot after $f_i$ finishes, it finds versions 5 and 8, but it will add version 5 because it is the most recent version smaller than 8, which is the $f_j$ executing version number.

The execution of a functionality $f$ in a transactional causal consistent context follows the following steps:
\begin{enumerate}
    \item When the functionality $f$ starts, $f.version = max(F.success.version) +1$, where $F.success.version$ is the set of version numbers of all the functionalities executions that finished successfully;
    \item Whenever an aggregate is read, it is selected a version according to do the functionality snapshot conditions, and added to it, if not already there;
    \item Whenever an aggregate is written, it is selected a version according to do the functionality snapshot conditions, and added to it, if not already there. In case the aggregate is being created, it is assigned the functionality version number. $f.written$ denotes the subset of $f.snapshot$ of the written aggregates;
    \item When the functionality finishes execution it proceeds to commit, and if the process succeeds the functionality finishes successfully, otherwise it aborts. It does the following actions:
    \begin{enumerate}
        \item The written aggregates preserve the intra-invariants, i.e. $\forall_{a \in f.written,\ irv \in a.intraInv} \text{ } a.irv$, where $a.intraInv$ denotes the aggregate intra invariants, and $a.irv$ denotes evaluation of the invariant in the $a$ version;
        \item For each aggregate versions to be written get, if exists, the most recent version of the same aggregate that was created by a concurrent functionality, i.e. $\forall_{a_i \in f.written}\ a_i.toMerge = max_{a.version}\{a_j \in a_i.versions: a_j.version \geq a_i.version\}$;
        \item For each aggregate that has a version to merge, merge it. $merge(a_i,a_i.toMerge)$ denotes the merged version; 
        \item Update the functionality version number to the most recent successfully finished functionality version number plus one ($f.version = max(F.success.version) +1$), and commit the written aggregates using the new functionality version number, $f.version$.
    \end{enumerate}
\end{enumerate}

Transactional causal consistency has the lost update anomaly, which occurs when there are several concurrently executing functionalities updating the same aggregates. The merge tries to handle this anomaly by verifying if it is possible to merge concurrent aggregate versions, while preserving the aggregate semantics and the intentions of the functionalities that changed them. 

The merge occurs in two versions of an aggregate, $merge(a_i,a_j)$, where $a_i$ is the version of the functionality to commit and $a_j$ the version already committed. To do the merge, it is necessary to find the version that is the common ancestor of both versions, in order to identify the differences. Since the version to commit evolved from a committed version, this version is the common ancestor and is denoted by $a_i.prev$. Note that between $a_i.prev$ and $a_j$ can be several other committed versions, if several concurrent functionalities have committed. Nevertheless, it is possible to identify which attributes were changed when compared with the common ancestor, which is denoted, respectively, by $diff(a_i.prev,a_i) \subseteq a.attributes$ and $diff(a_i.prev,a_j) \subseteq a.attributes$, where $a.attributes$ denotes the attributes of $a.aggregate$. Consider the example above, where, instead of two, there are three concurrent functionalities executing with version number 8, $f_i$, $f_j$ and $f_k$, and all writing aggregate $a$. Suppose that $f_i$ finishes first with version number 8, and then $f_j$ with version number 9, when $f_k$ tries to finish will find aggregate $a$ with version 9. The common ancestor will be version 5, and so the differences will be between version 5 and 9 and version 5 and the version $f_k$ is trying to write.

The aggregate designer has to define the semantics of consistent merges, which depends on the aggregate semantics. The idea is that the merge should preserve the intention of each one of the functionalities. For instance, suppose a functionality that changes the start and end dates of a tournament, and one user invokes the functionality to change the start date while keeping the end date, and another user, concurrently change the end date while preserving the start date. In this case it does not make sense to merge the two versions, because it would dismiss the intention of each one of the users, they changed one of the dates while observing the value of the other one. So, the merge can only occur if the merge does not violated the intention of each one of the functionalities. Note that, the intention semantics is not an intra-invariant, it has to do with TCC semantics. Additionally, if both functionalities changed all the attributes of an intention, we cannot say that a functionality violates the intention of the other.

Therefore, given an aggregate $a$, the subsets of attributes that cannot be simultaneously changed in different versions are denoted by $a.intentions \subseteq \mathcal{P} (a.attributes)$. The following condition defines the cases where the merge cannot occur, the lost update anomaly cannot be handled, and the functionality has to abort:
\begin{equation*}
\begin{split}
    & \exists_{at_i \in diff(a_i.prev,a_i), at_j \in diff(a_i.prev,a_j), at_i \neq at_j} \exists_{i \in a_i.intentions}: \\
    & \{at_i,at_j\} \subseteq i\ \land \\ & \neg (i \subseteq diff(a_i.prev,a_i) \land i \subseteq diff(a_i.prev,a_j))
\end{split}
\end{equation*}

The second step of the merge process is to merge any changed attributes, which is done using predefined merge methods defined by the developer. Note that, it may not be possible to merge attributes, in which case the merge fails, which is similar to the of non-incremental operations concept defined in the conflict-free data types literature. Therefore, the merge is defined using developer defined $merge$ methods per attribute:
\begin{equation*}
\begin{split}
    at \in diff(a_i.prev,a_i) \cap diff(a_i.prev,a_j) \\ \implies merge(a_i,a_j,at)
\end{split}
\end{equation*}

Overall, the process of merging two aggregate versions follows the steps:

\begin{enumerate}
    \item Verify intention conditions;
    \item Merge the changed attributes;
    \item Run intra-invariants in the merged version.
\end{enumerate}

\begin{minipage}{\linewidth}
\begin{lstlisting}[language=Java, basicstyle=\tiny, caption=Extend Aggregate with Merge Semantics, label={lst:mergeSemantics}]

Tournament Causal Consistency
    Merge 
        Intentions Tournament t1, t2:
            (t1.startTime, t2.endTime)
            ...
        Methods Tournament tN, tC:
            startTime
                this.startTime = tN.startTime
    ...
\end{lstlisting}
\end{minipage} 

The specification in Listing~\ref{lst:mergeSemantics} illustrates the extension of the \textit{Tournament} aggregate for transactional causal consistency, where the merge method for the start date uses the most recent update (\textit{tN} is the new version whereas \textit{tC} is the committed). If the intention is not violated, overwrite is allowed, the last functionality to commit overwrites the start date.

\subsection{Eventual Consistency}

The transactional model for aggregates has an additional level of complexity, due to the upstream-downstream relation between aggregates. A functionality cannot directly interact with aggregates that are downstream of its main aggregate. So, if the execution of the functionality has some impact on downstream aggregates it cannot occur in the context of the causal consistent transaction executing the functionality. Instead, an event is published and eventually processed by the interested subscribing downstream aggregates. These events are inferred from the downstream inter-invariants.

\begin{minipage}{\linewidth}
\begin{lstlisting}[language=Java, basicstyle=\tiny, caption=Extend Aggregates with Events, label={lst:eventsExtension}]

CourseExecution Causal Consistency
    Events
        Publish
            ANONYMIZE_STUDENT
            
Tournament Causal Consistency
    Events 
        Subscribe
            ANONYMIZE_STUDENT: PARTICIPANT_EXISTS, ...
\end{lstlisting}
\end{minipage} 

Therefore, it is necessary to extend the aggregate specification with this situation. Listing~\ref{lst:eventsExtension} shows the causal consistency extension due to the tournament \textit{PARTICIPANT\_EXISTS} inter-invariant. The processing of the event triggers the execution of functionality \textit{anonymizeParticipant} in Listing~\ref{lst:tournamentFunctionalities}. In the example, a version of tournament subscribe to the \textit{ANONYMIZE\_STUDENT} events emitted by the course execution version inferred from the \textit{Participant is} declaration, where participant \textit{id from s.id}, see Listing~\ref{lst:tournamentAggregate}. The set of events an aggregate version subscribes to is dynamically calculated and it may depend on the aggregate version information. In the example, a tournament subscribes the anonymize events that refer to its participants, which can change dynamically through, for instance, the add participant functionality.

Due to the occurrence of these events, it is necessary to reassess the causal consistency associated to the functionality execution. The causal snapshot associated with functionality execution should also guarantee that all the aggregates subscribing to an event are consistent when involved in a causal transaction, i.e. they have processed all the events in common.

As an example, consider three different aggregate versions $a_i$, $a_j$ and $a_k$, where the last two subscribe events of the first one. In a causal snapshot two cases can occur: (1) $a_j$ and $a_k$ belong to the snapshot but $a_i$ does not; (2) $a_i$ and $a_j$ belong to the snapshot. Other cases are similar or combinations of these two cases. In the first case, it is necessary to guarantee that both versions process the same events they are subscribed to. In the second case, it is necessary to guarantee that all the events emitted by $a_i$ have been processed by $a_j$, if it subscribes them.

Given an event $e \in E$, where $E$ is the set of all events, $e.type$ denotes the type of event, $e.aggregate$ the aggregate that emits the event, and $e.version$ denotes the version of the aggregate emitting the event. Given an aggregate version $a$, $a.subsEvs \subseteq E$ denotes the events it subscribes to, and $a.emitEvs \subseteq E$ the set of events it has emitted. 

An aggregate event subscription is a condition that uses the event sender aggregate version and the event type to identify the subscribed events. The sender aggregate version indicates the last version of the aggregate sender the aggregate subscriber depends on. Therefore, after an aggregate processes an event it updates the event sender aggregate version. For instance, tournament subscribes the anonymize event of course execution, it contains the course execution version it subscribes to. Note that, it is not necessarily the most recent version of course execution, because the course execution may have evolved without sending events relevant for the tournament. Of course, it can also be the case that the course execution emitted an event the course execution subscribed to but has not processed it yet. Therefore, the subscriber can process events sent by aggregates that have a higher version number than the version it subscribes to. This is evaluated by the subscription condition. We are also considering that the publish-subscribe channel is causal order. 

Therefore, to handle the impact of eventual consistency in the system, two additional conditions have to be added to to the functionality causal snapshot.

For the first condition, if two aggregates subscribe to an event of an upstream aggregate and they are in causal snapshot, they should be consistent according to the processing of the event, even if the upstream aggregate is not part of the causal snapshot. 
\begin{equation*}
\begin{split}
    & \forall_{a_j,a_k \in f.snapshot, a_i \in A} \forall_{e \in a_i.emitEvs} \\
    & e \in a_j.subsEvs \cap a_k.subsEvs \text{ } \lor \\
    & e \notin a_j.subsEvs \cup a_k.subsEvs 
\end{split}
\end{equation*}

Therefore, either both subscribe the event, and have not processed it, or do not subscribe it, which means that is already processed or is not relevant.

The second condition specifies that if upstream and downstream aggregates are in the snapshot, all the events emitted by the former have to be processed by the latter.
\begin{equation*}
\begin{split}
    & \forall_{a_i,a_j \in f.snapshot} \forall_{e \in a_i.emitEvs} \\
    & e \notin a_j.subsEvs
\end{split}
\end{equation*}

If the emitted event does not belong to $a_j$'s subscribed events, either it was already processed or is not relevant for $a_j$. 

Events are emitted at functionality commit time, to guarantee atomicity of the creation of the new aggregate version and the emission of the event. The event version is equal to the functionality version that did the change in the aggregate, which is the aggregate version as well. Similarly, an event is considered processed when the functionality it triggered commits. An event may be associated with more than one inter-invariant, if it depends on the same change of the upstream aggregate, which corresponds, actually, to the first case above.

\section{Simulator}
\label{sec:simulator}

To experiment the design and behaviour of functionalities business logic execution, using transactional causal consistency, on a microservice system implemented as a set of upstream-downstream aggregates, we developed a simulator that replicates the relevant aspects of aggregate executing under TCC. 

\subsection{Architecture}

The simulator uses an architecture where each functionality is available as a web-service, the interactions with the main and upstream aggregates is done using transactional invocations and the interactions with downstream aggregates is done writing events in a database that is periodically queried by a scheduler. When the event is detected by the scheduler it is also processed using transactional causal consistency. The \textit{Unit of Work} pattern~\cite[Chapter~11]{fowler03} is used for implementation of transactional causal consistency.

\begin{figure}[htbp]
\centerline{\includegraphics[width=7cm]{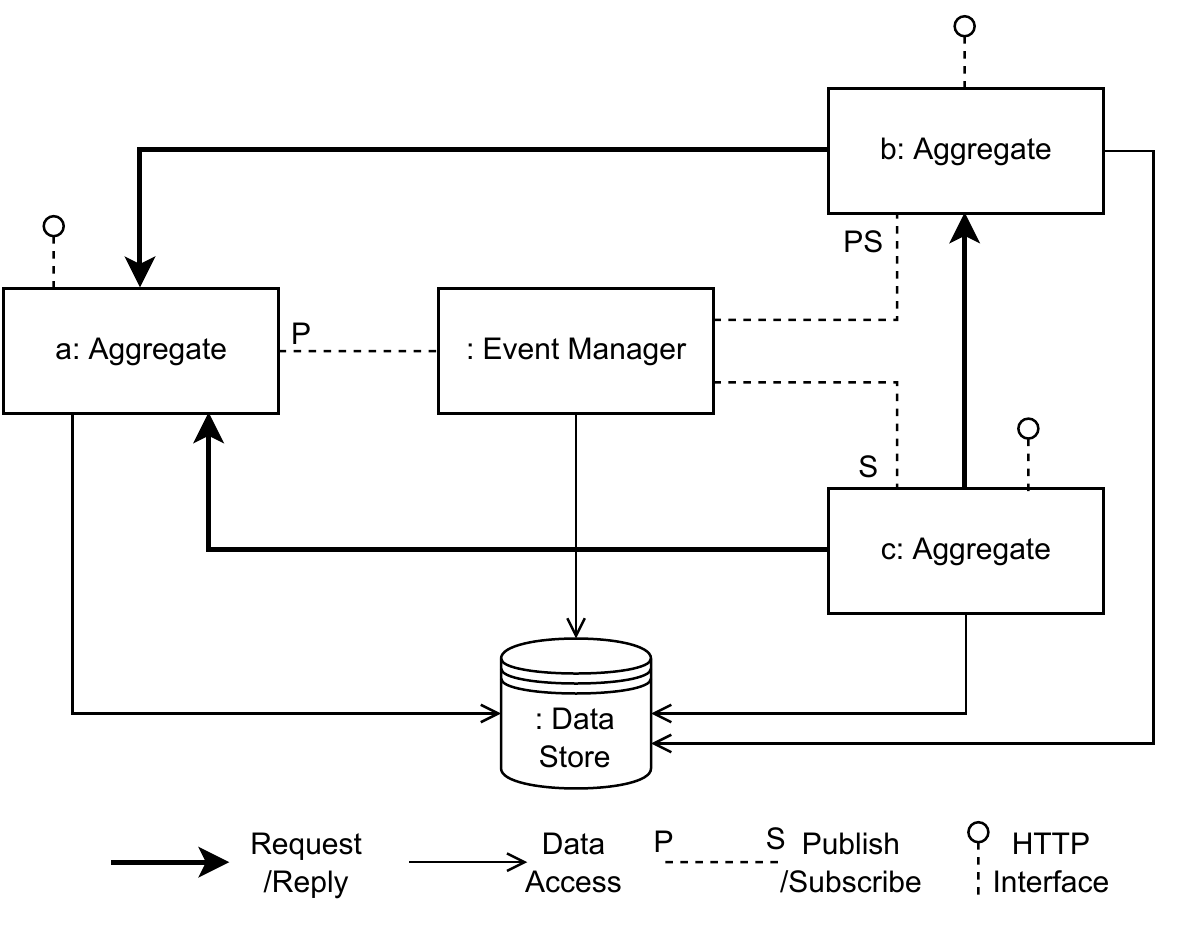}}
\caption{Simulator: Component-and-Connector View}
\label{fig:component-and-connector}
\end{figure}

Figure~\ref{fig:component-and-connector} presents a component-and-connector view~\cite{clements11} of the simulator architecture. It has three component types, \textit{Aggregate}, \textit{Event Manager} and \textit{Data Store}, and four connector types. The request/reply connector implements the upstream-downstream relations between aggregate components, where the invoker is the downstream aggregate. The publish-subscribe connector implements the event channels, which are managed by the \textit{Event Manager} component, and have publish and subscribe roles, for instance, aggregate \textit{b} is both a publish and subscriber because it emits events that are consumed by its downstream aggregate component \textit{c} and subscribes events from its upstream aggregate \textit{a}. Aggregate components use data access connectors to manage their aggregates persistence, whereas the event manager component used the data access connector to persist the events. Finally, the aggregate components have a HTTP interface connector that it is used to start its functionalities, the functionalities that have the component aggregate as their main aggregate.

\begin{figure}[htbp]
\centerline{\includegraphics[width=7cm]{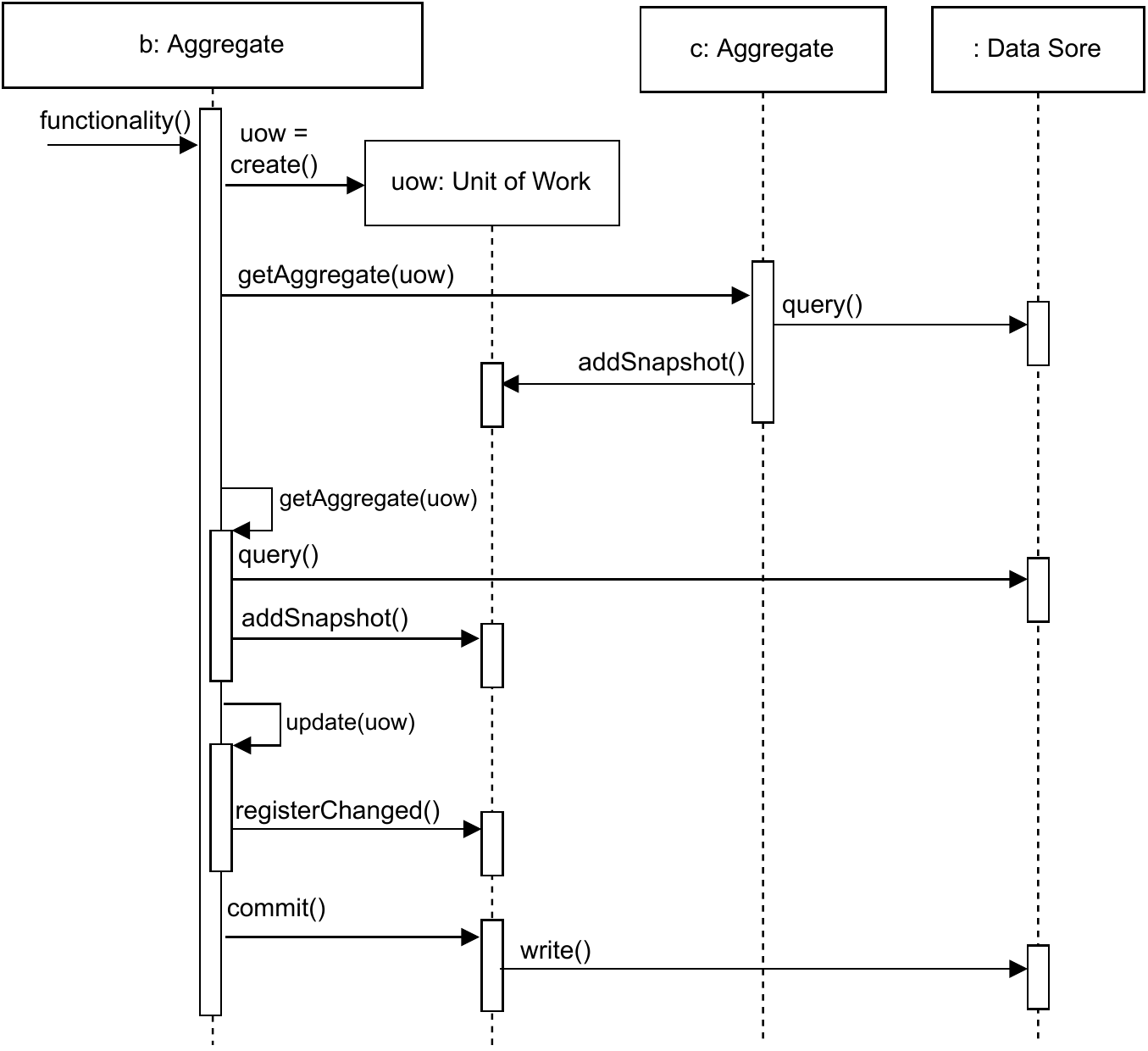}}
\caption{Simulator: Functionality Interaction}
\label{fig:interaction-diagram}
\end{figure}

Figure~\ref{fig:interaction-diagram} depicts the interactions associated with the execution of a simple functionality of aggregate \textit{b}, which reads aggregate \textit{c}. The execution, which occurs in the component of the main aggregate, starts by creating a unit of work, that is responsible to manage the causal transaction. The read of aggregates is intercepted by the unit of work, that adds the version to the causal snapshot. The changes done on the \textit{b} aggregate are registered in the unit of work. On commit, the unit of work does the verifications and required merges, finishing by writing the new version of the aggregate and the events to emit in the data store.

The simulator exercises the causal transactional behavior of the functionality. Each aggregate invocation is done in a system transaction, which allows the interleaving with other executing functionalities. The commit is a serializable transaction, which guarantees the atomic write of all versions, where each version is written with the unit of work version number. On the other hand, the events are also written in the commit, which guarantees that their emission, and processing, is atomic with the aggregates changes.

The processing of events is periodically triggered by the event manager, that checks the events in the data store and start their processing in their own functionality, which executes in a causal transaction transaction. Therefore, during a functionality execution it is possible that one of its aggregates to update is changed in the consequence of the processing of an event, which will be detected during the commit and can trigger a merge.

\subsection{Implementation}

The simulator is implemented as a Java Spring Boot web application, where the aggregate functionalities are published as web services, and aggregates are accessed through Spring transactional services. The aggregates are stored in a PostgreSQL database accessed using Hibernate. The emitted events are also stored in the database.

\begin{figure}[htbp]
\centerline{\includegraphics[width=7.5cm]{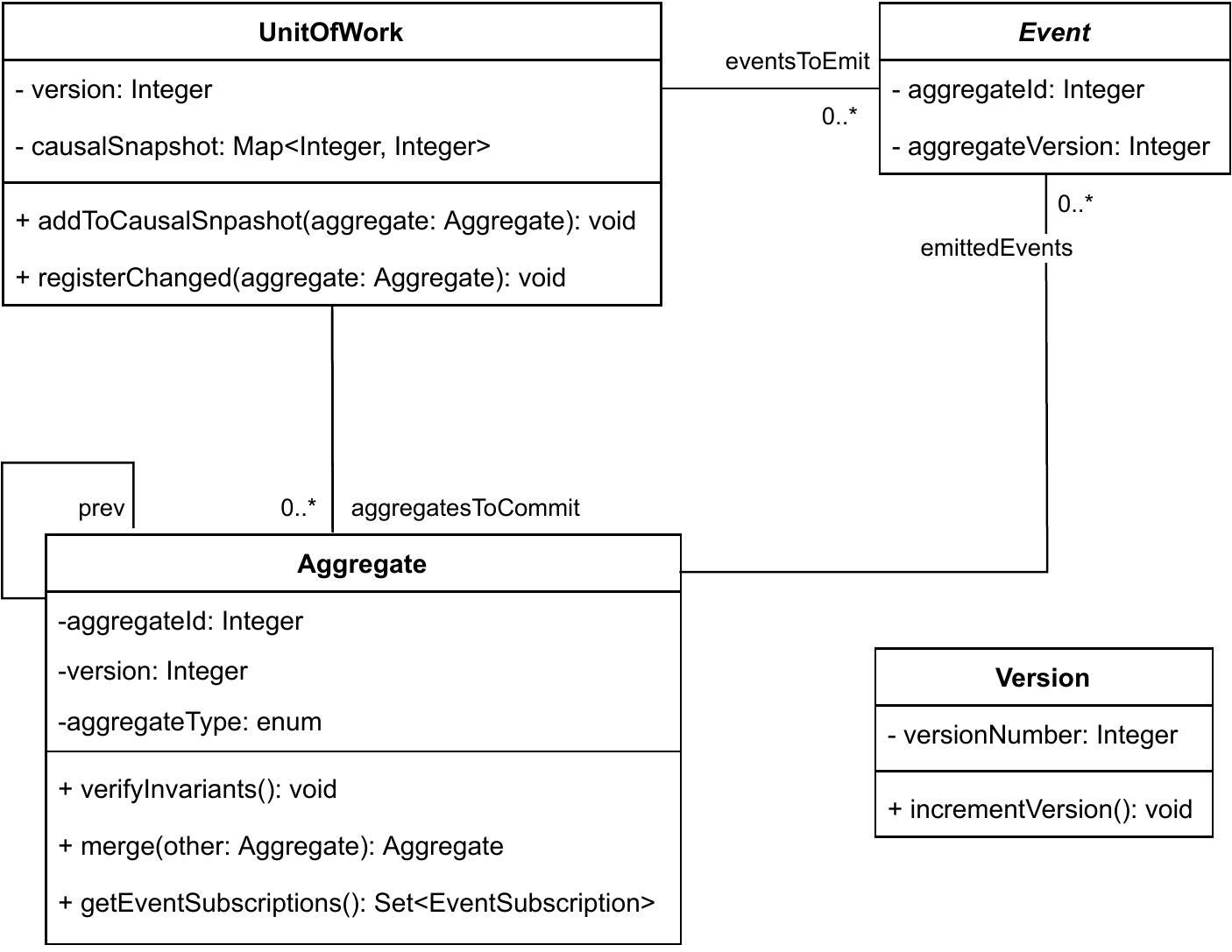}}
\caption{Simulator: Domain Model View}
\label{fig:domain-model-diagram}
\end{figure}

Figure~\ref{fig:domain-model-diagram} presents the simulator domain model, which is the core of the simulator and it is encapsulated by services. \textit{Version} is a persistent singleton entity that contains the version number of most recent committed causal transaction. \textit{Event} contains the identification and version of the aggregate that emit the event. Events are stored in a database and are periodically checked by event handlers, which trigger the handling. The \textit{Aggregate} contains its identification, version, and aggregate type. The \textit{aggregateType} refers to the type of aggregate, which is set in subclasses. Aggregates are implemented as a group of embedded elements such that can be read and written to the database as a whole. Additionally, although all aggregates are implemented in the same database, they are completely independent, not having any referential dependency. Having all aggregates in the same database simplifies the unit of work atomic commit. The aggregate refers to its previous version, that is used by the merge method. On the other hand, each aggregate version has the set of events it emitted. The three aggregate methods in Figure~\ref{fig:domain-model-diagram} are extended in the subclasses with aggregate-specific business logic. The \textit{Unit of Work} is a non persistent entity that contains the version of the causal transaction executing the functionality and the causal snapshot. While being changed in unit of work, attribute reference \textit{aggegatesToCommit}, aggregates are detached from the database, which creates a version whose changes are local to the functionality execution. When committing, the unit of work creates new aggregate versions with its version number and store them in the database. Additionally, the commit stores the \textit{emittedEvents} in the database, which corresponds to its emission and processing. The unit of work commit is implemented in the service, not in the domain entity, because Spring provides transactional behavior at the service level.

\subsection{Extension}

To simulate a microservice system using transactional causal consistency, the developer has to extend from the implementation abstract classes. 

\begin{lstlisting}[language=Java, basicstyle=\tiny, caption=Tournament Aggregate Implementation, label={lst:tournamentImplementation}]

@Entity
public class Tournament extends Aggregate {
    private LocalDateTime startTime;
    private LocalDateTime endTime;
    @ElementCollection
    private Set<TournamentParticipant> participants;

    @Override
    public Set<EventSubscription> getEventSubscriptions() {
        Set<EventSubscription> eventSubscriptions = new HashSet<>();
        if (this.getState() == ACTIVE) {
            interInvariantParticipantExists(eventSubscriptions);
            ...
        }
        return eventSubscriptions;
    }
    private void interInvariantParticipantExists(
                        Set<EventSubscription> eventSubscriptions) {
        eventSubscriptions.add(
            new EventSubscription(this.courseExecution.getAggregateId(), 
            this.courseExecution.getVersion(), UNENROLL_STUDENT, this));
        eventSubscriptions.add(
            new EventSubscription(this.courseExecution.getAggregateId(), 
            this.courseExecution.getVersion(), ANONYMIZE_EXECUTION_STUDENT, this));
    }     
    @Override
    public void verifyInvariants() {
        if (!(invariantAnswerBeforeStart() && ...)) {
            throw new TutorException(INVARIANT_BREAK, getAggregateId());
        }
    }

    public boolean invariantStartTimeBeforeEndTime() {
        return this.startTime.isBefore(this.endTime);
    }

    @Override
    public Set<String> getFieldsChangedByFunctionalities()  {
        return Set.of("startTime", "endTime", 
                      "participants", ...);
    }

    @Override
    public Set<String[]> getIntentions() {
        return Set.of(
                new String[]{"startTime", "endTime"},
                ...);
    }

    @Override
    public Aggregate mergeFields(Set<String> toCommitVersionChangedFields, 
        Aggregate committedVersion, Set<String> committedVersionChangedFields ){
        if (!(committedVersion instanceof Tournament)) {
            throw new TutorException(AGGREGATE_MERGE_FAILURE, getAggregateId());
        }

        Tournament committedTournament = (Tournament) committedVersion;
        Tournament mergedTournament = new Tournament(this);

        mergeStartTime(toCommitVersionChangedFields, committedTournament, 
            mergedTournament);
        ...

        return mergedTournament;
    }

    private void mergeStartTime(Set<String> toCommitVersionChangedFields, 
        Tournament committedTournament, Tournament mergedTournament) {
        if (toCommitVersionChangedFields.contains("startTime")) {
            mergedTournament.setStartTime(getStartTime());
        } else {
            mergedTournament.setStartTime(committedTournament.getStartTime());
        }
    }
    ...
}
\end{lstlisting}

Listing~\ref{lst:tournamentImplementation} shows the implementation of the tournament aggregate excerpt previously specified. Attribute \textit{participants} is an embedded collection, which is loaded with the aggregate. The aggregate declares the events it subscribes, by overriding \textit{getEventSubscriptions}, which are defined for each of the inter-invariants, \textit{interInvariantParticipantExists}. It also defines its intra-invariants and include their invocation in the override \textit{verifyInvariants} method, which is invoked during commit. Additionally, it is only necessary to declare what are the attributes changed by functionalities, \textit{getFieldsChangedByFunctionalities}, the intentions, \textit{getIntentions}, and the attributes merge functions, \textit{mergeFields}, to feed the merge method that is defined in the \textit{Aggregate} abstract class. In the case of start time, if a merge is possible, and the functionality changed it, it is set to the new value, keeping the old value, otherwise, as described by method \textit{mergeStartTime}. As can be observed, a minimal set of methods need to be defined in each subclass to the define the aggregate business logic.

Listing~\ref{lst:tournamentFuncImpl} illustrates the functionality for updating a tournament. How it creates the unit of work, pass it in the invocations of transactional services, and finishes committing it. The update service uses the unit of work to create the causal snapshot and register the updated aggregates. The method \textit{getCausalTournamentLocal} does the query to obtain the correct tournament version and add it to causal snapshot.

\begin{lstlisting}[language=Java, basicstyle=\tiny, caption=Tournament Functionalities Implementation , label={lst:tournamentFuncImpl}]

public class TournamentFunctionalities {
    public void updateTournament(TournamentDto tournamentDto, ...) {
        UnitOfWork unitOfWork = unitOfWorkService.createUnitOfWork();
        ...
        TournamentDto newTournamentDto = 
        tournamentService.updateTournament(tournamentDto, ..., unitOfWork);
        ...
        unitOfWorkService.commit(unitOfWork);
}

@Service
public class TournamentService {
    @Transactional
    public TournamentDto updateTournament(TournamentDto tournamentDto,
                                          ... , UnitOfWork unitOfWork) {
        Tournament oldTournament =  
            getCausalTournamentLocal(tournamentDto.getAggregateId(), unitOfWork);
        Tournament newTournament = new Tournament(oldTournament);

        if (tournamentDto.getStartTime() != null ) {
            newTournament.setStartTime(
                LocalDateTime.parse(tournamentDto.getStartTime()));
            unitOfWork.registerChanged(newTournament);
        }

        if (tournamentDto.getEndTime() != null ) {
            newTournament.setEndTime(
                LocalDateTime.parse(tournamentDto.getEndTime()));
            unitOfWork.registerChanged(newTournament);
        }
        ...
        return new TournamentDto(newTournament);
    }
    ...
}
\end{lstlisting}

\section{Case Study}
\label{sec:caseStudy}

QuizzesTutor\footnote{\url{https://quizzes-tutor.tecnico.ulisboa.pt/}} 
is a large monolith application for teachers to prepare different types of questions and propose quizzes to students. Students can answers quizzes, generate quizzes for self-assessment, selecting the questions topics, and organized quizzes tournaments, among other features. It is a business logic rich system implemented as a web application.

\begin{figure}[htbp]
\centerline{\includegraphics[width=6cm]{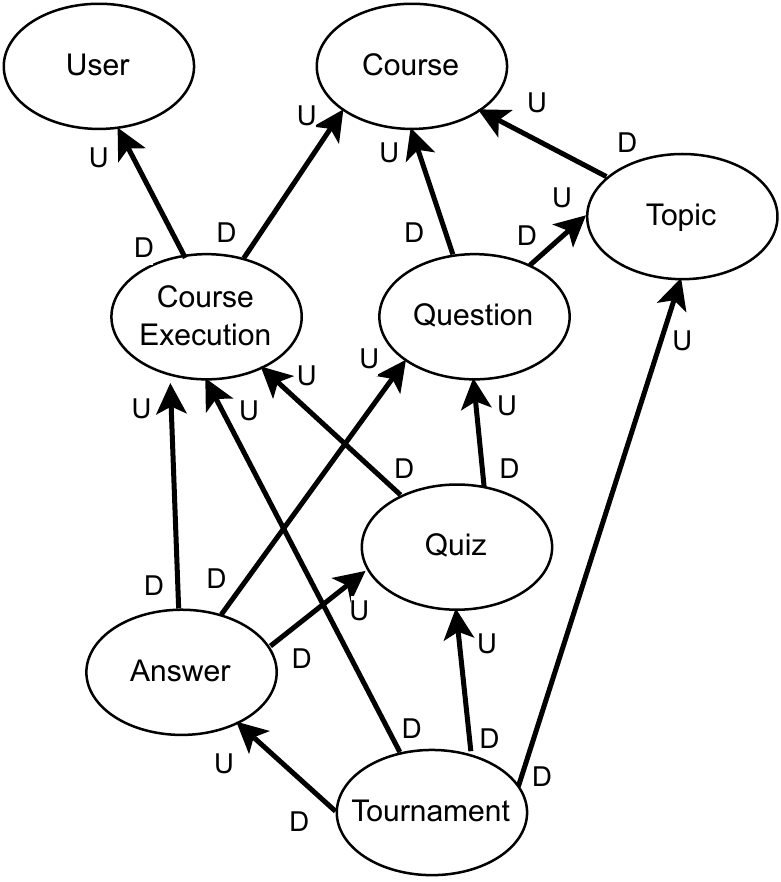}}
\caption{QuizzesTutor Inter-Aggregate Context Map}
\label{fig:aggregate-relations}
\end{figure}

To experiment the design of business logic using transactional causal consistent aggregates, a set of aggregates and their upstream-downstream relations were identified in a subset of the QuizzesTutor system, see Figure~\ref{fig:aggregate-relations}. 

A \textit{Course} is a course in a teaching institution. A \textit{Course Execution} corresponds to an edition of a \textit{Course} in a given year. A \textit{User} represents a person using the system and can be of one of the following 3 types: \textit{STUDENT}, \textit{TEACHER} and \textit{Admin}. A \textit{Student} can enroll himself in \textit{Course Executions}, a \textit{Teacher} can create \textit{Questions} in a \textit{Course}, which can be use in all its course executions to create quizzes. An \textit{ADMIN} is responsible for creating, updating and deleting courses and \textit{Course Execution}. A \textit{Question} has associated \textit{Topics}. A \textit{Quiz} contains a set of questions for students to answer. An \textit{Answer} contains the student's answer to a quiz. A \textit{Tournament} is created by a student for students to answer in the context of a competition. Tournaments have an associated quiz whose questions are randomly selected, given the tournament topics.

Figure \ref{fig:aggregate-relations} shows the aggregates and the upstream-downstream relations between them.

\section{Evaluation}
\label{sec:evaluation}

\subsection{Correctness}

To verify that the simulator correctly handles situations where the concurrency anomalies can occur, addressing the second research question, we designed scenarios that exercise the interleaved execution of functionalities and event handling. For each of the scenarios, JMeter\footnote{\url{https://jmeter.apache.org/}} tests were also written. 

JMeter allows for the execution of thread groups concurrently. However it is not possible to control how they interleave. In order to have better control over concurrent execution of functionalities, we have decided to run all threads groups sequentially and simulate concurrency by manipulating the version number of the system. A functionality which writes an aggregate finishes and commits with version number $n$. When another functionality starts, it has version number $n+1$, which means if it reads the aggregate versions created by first functionality. This is a standard sequential execution. If we want to simulate a concurrent execution of these two functionalities, we can decrement the system version number by 1, between their executions. This results in the second functionality also starting with a version number of $n$ like the first, instead of $n+1$. This ensures that the second functionality reads the same aggregate versions read by the first functionality. If the second functionality also performs a write on the same aggregate, upon committing it detects the version written by the first functionality as concurrent, initiating a merge operation between the two. We can decide which functionality does the merge by running it after the version decrement. After the concurrent functionalities are executed, the system version number is incremented by the previously decremented amount to guarantee that further actions observe the state set by the the previously finished functionality. The system version number increment and decrement is done by a SQL update on the respective table, through the JMeter API.

Another technique was used to handle event detection. Event detection and processing runs periodically on a predetermined interval, which depends on when the last event handling finished. However, during tests, sometimes it is useful to trigger the event detection at a precise moment in the test, for instance when trying to test the concurrent execution of an event handling and a functionality. To achieve it, we created web services that trigger the detection and processing of a specific event type for an aggregate. On the other hand, to also test the periodic event detection, if the test behavior allows it, we use a constant timer that stops the test for a few milliseconds, waiting for the event processing to occur.

There are situations where the periodic event detection, which occurs every second, can compromise the expected test behavior. Therefore, another request is implemented, to be used by the JMeter tests, that disable and enable the periodic event detection whenever necessary.

\begin{figure*}
\centering
\subfloat[Sequential: Update, Add, Event\label{fig:concurrency-interleaving-1}]{\hspace{0.05\linewidth}\includegraphics[width=0.4\linewidth]{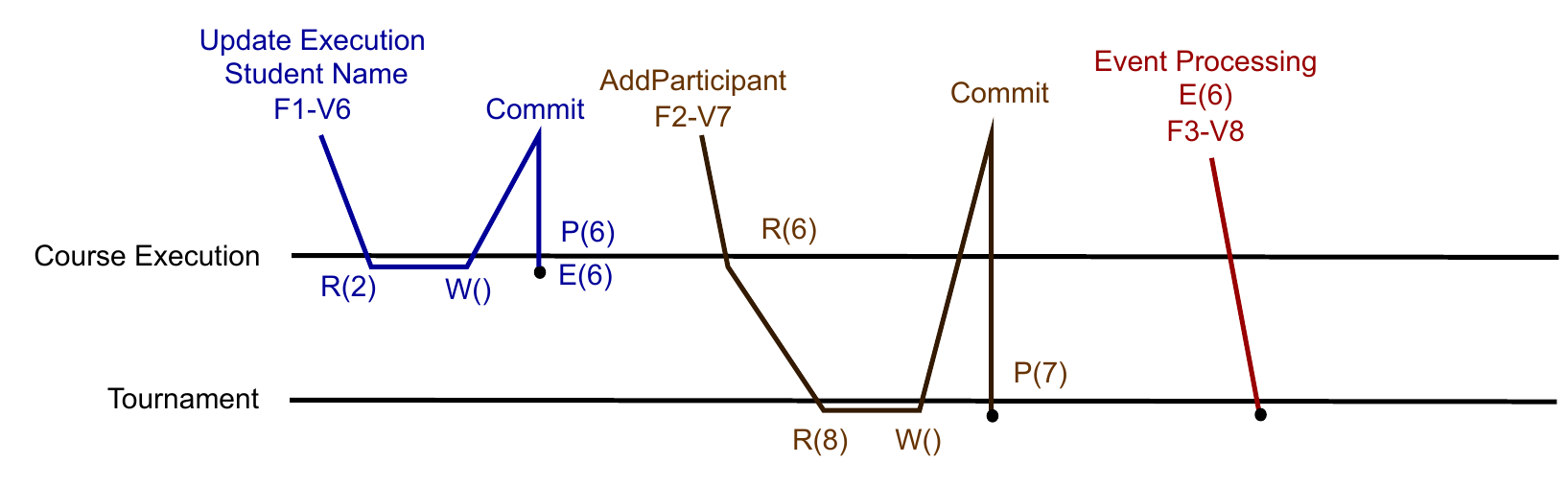}}
\subfloat[Sequential: Add, Update, Event\label{fig:concurrency-interleaving-2}]{\includegraphics[width=0.4\linewidth]{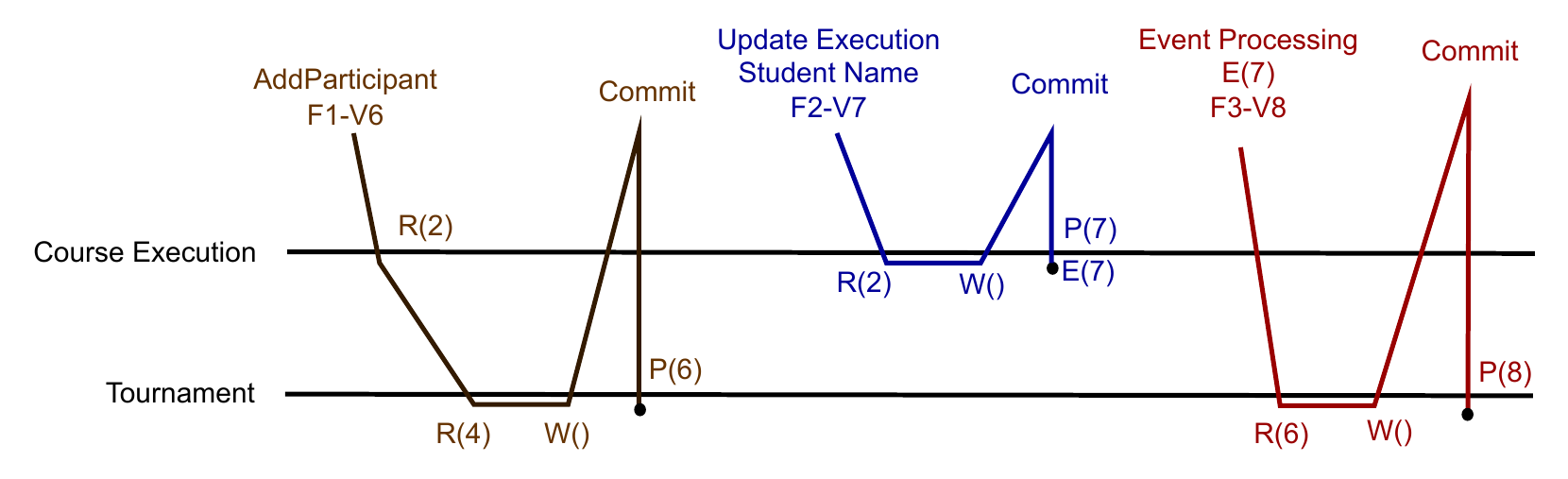}}
\newline
\subfloat[Concurrent: Update(1), Add, Update(2), Event\label{fig:concurrency-interleaving-3}]{\includegraphics[width=0.4\linewidth]{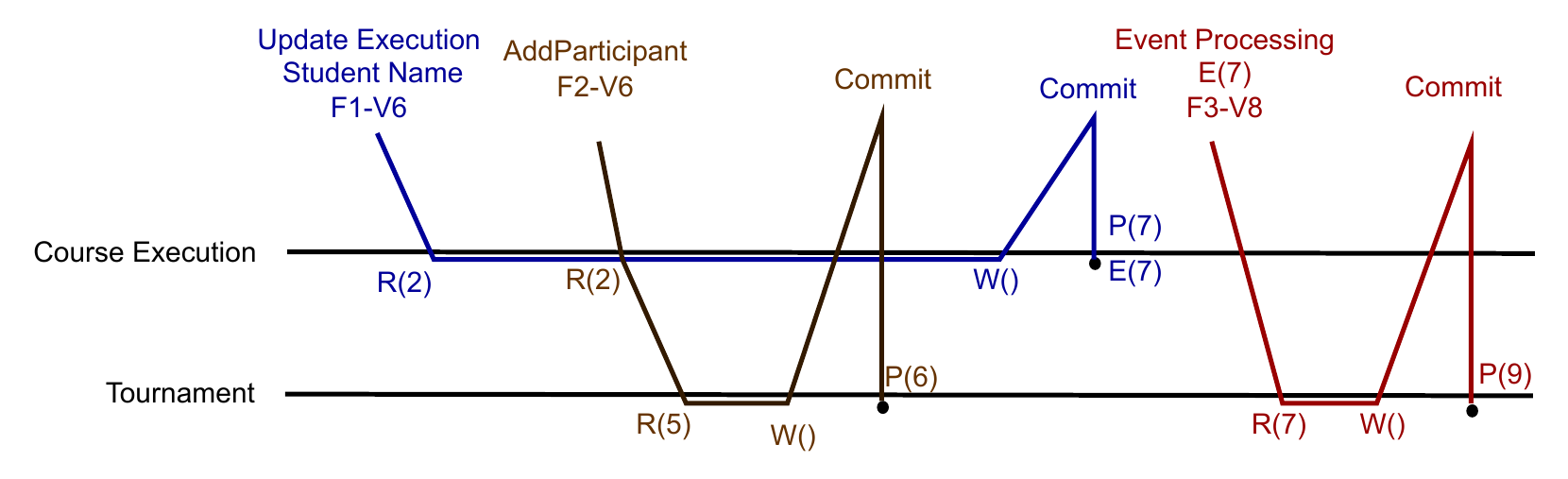}}
\subfloat[c\label{fig:concurrency-interleaving-4}]{\includegraphics[width=0.4\linewidth]{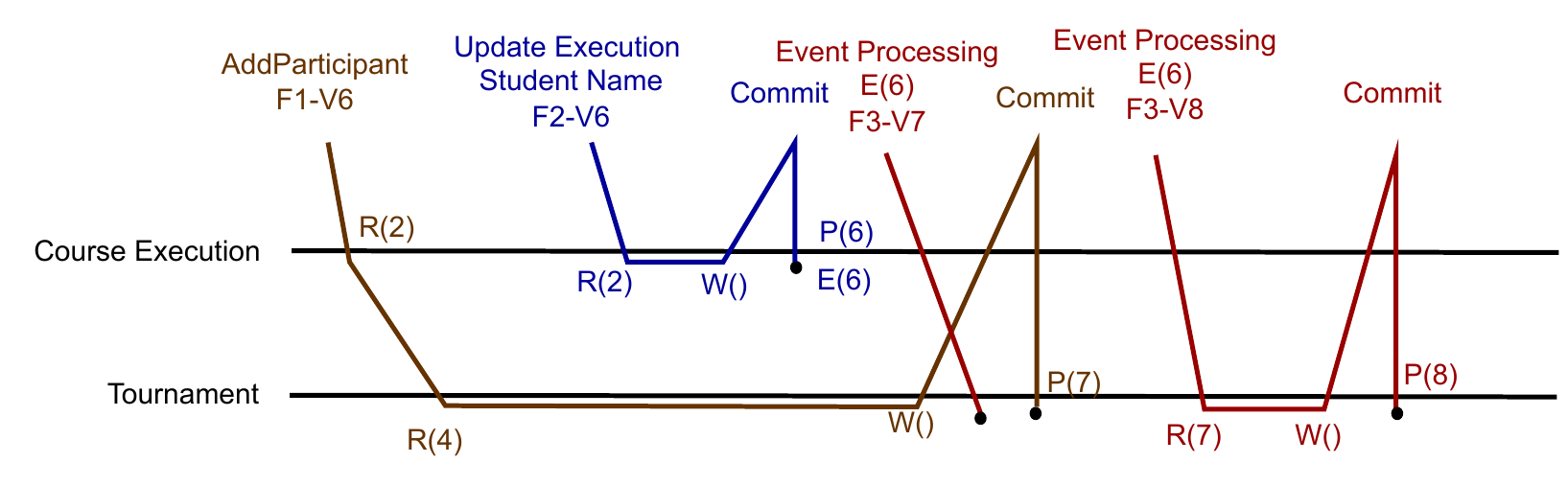}}
\caption{Functionalities and Event Processing Interleaving}
\label{fig:concurrency-interleavings}
\end{figure*}

In the scenarios, R, W, P and E, denote, respectively, the read of a version, write in a version inside the unit of work, write a new version (persistent), and emit an event associated with a new version. The functionality and event processing versions correspond to their initial version number. Commits and merges are operations that can span several aggregates and are represented with a vertical line to represent their atomicity. Actual, commit and merge belong to the same atomic transaction, but they are represented by different vertical lines for clarity.

Figure~\ref{fig:concurrency-interleavings} shows the different interleaving associated with the execution of update student and add participant functionalities, where the former emits an event. 
In Figure~\ref{fig:concurrency-interleaving-1} add participant succeeds when it adds the tournament to the causal snapshot, because the event emitted by course execution is not subscribed by tournament. This occurs because the updated student is not a tournament participant. Afterwards, the event is not detected by the tournament because it already contains the updated participant.
In figure~\ref{fig:concurrency-interleaving-2} the add participant commits first and the event processing occurs after both functionalities commit. The event processing changes the participant with the new information, because the tournament subscribes to changes in the course execution version, 2 to 7, where the student name changed.
In Figure \ref{fig:concurrency-interleaving-3}, illustrates the case when both functionalities execute concurrently and add participant finishes first. Like in the previous case, the event processing updates the participant with the new information.
Figure~\ref{fig:concurrency-interleaving-4} illustrates the other concurrent case, but the update student finishes first. In the scenario the event detection occurs twice. First it is not detected, because the tournament version 4 does not contain the student as participant. A second time after add participant commits, where tournament subscribes course execution versions higher than 2, the event is version 6, and the changed student is a participant of tournament version 7.

\begin{figure}[htbp]
\centerline{\includegraphics[width=8cm]{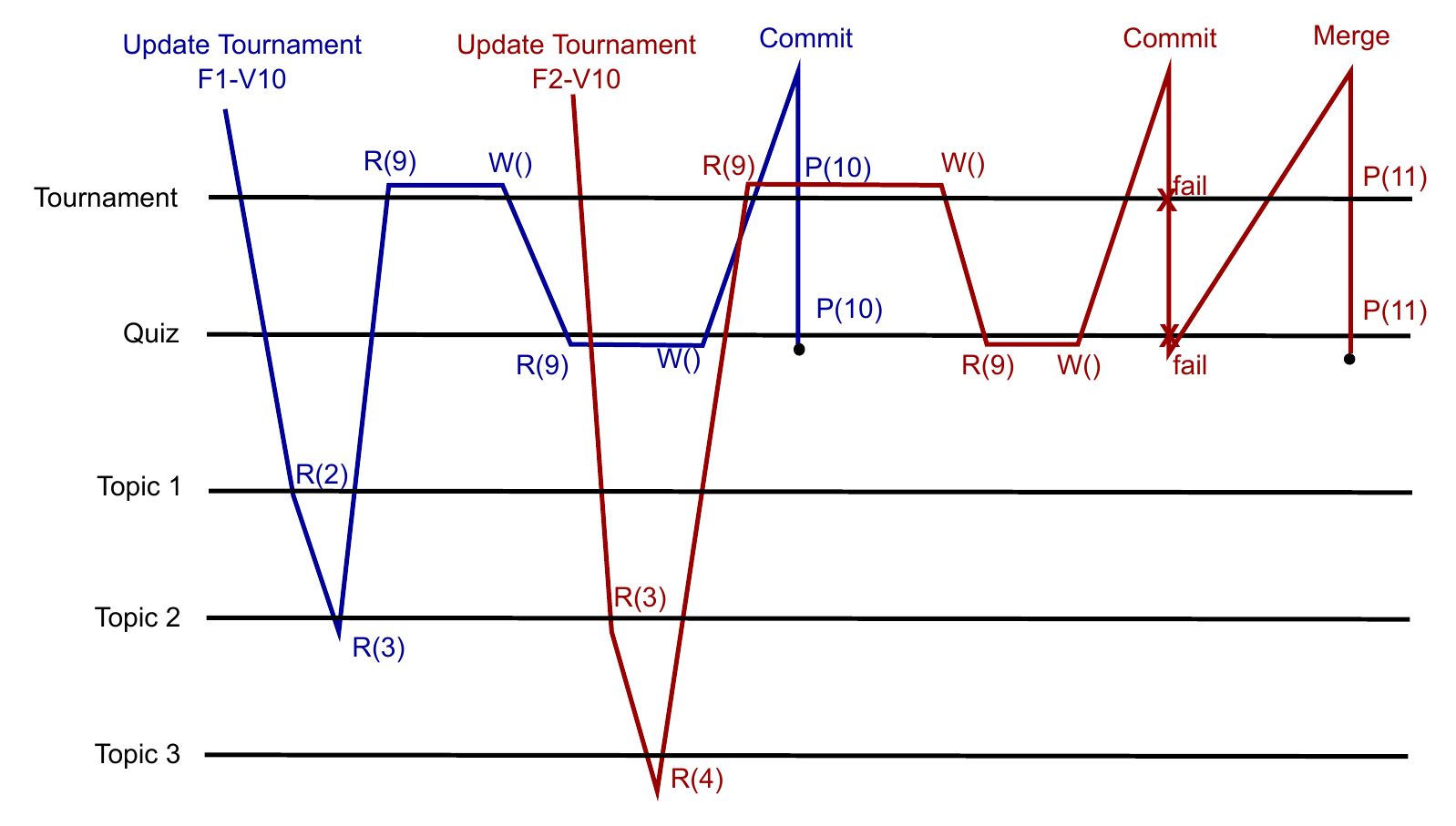}}
\caption{Concurrent complex functionalities}
\label{fig:concurrency-interleaving-5}
\end{figure}

Figure \ref{fig:concurrency-interleaving-5} represents the concurrent execution of two complex functionalities, the update of tournament. The first commits without any problems. When the second tries to commit it encounters a concurrent tournament and a concurrent quiz. Both have to be merged. In this case, the merge methods override the previous versions with the last versions, such that topics and quiz are consistent. 

\subsection{Experiment Analysis}

In the experiment we implemented the business logic of 42 functionalities: 14 queries; 24 simple functionalities; and 4 complex functionalities. The experiment allowed us to identify general guidelines for the business logic implementation and some particular cases, which provide useful insights.

Overall, due to the use of transactional causal consistency the implementation of the functionalities business logic is quiet uniform over all types of functionalities. Additionally, due to the causal snapshot, queries never abort. In what concerns simple and complex functionalities, it is necessary to be more careful in order to handle the lost update anomaly, which require the definition of intentions and operation merge methods. Anyway, and as it will be discussed in the complexity analysis subsection, the definition of the semantics can fortunately be done at the aggregate granularity level.

Although most of the implementations of inter-invariants result in business logic where the downstream aggregate has to handle the upstream aggregate events, this is not always the case. We have identified a situation where the downstream aggregates \textit{owns} the upstream aggregate. The upstream aggregate is mostly updated by downstream aggregate functionalities. Therefore, the change in the upstream aggregate does not need to be propagated to the downstream aggregate. It is up to the downstream functionality to guarantee the consistency, and this occurs in the context of transactional causal consistency functionality. In particular, the tournament creates a quiz, and, even though, the quiz is a upstream aggregate it is mostly changed by tournament functionalities. The tournament may still have to subscribe to quiz events but those events are not the result of a direct change to the quiz, instead they are result of events processed by the quiz, which it cannot control.


The merge between aggregates during commit is specified at the aggregate granularity. However, if the merge occurs in a complex functionality it is necessary to guarantee that all merges are consistent. Therefore, it may be necessary to consistently define the intentions and merge operations of the different aggregates. The functionality that updates a tournament also may update the quiz. The tournament start and end times are also attributes of the quiz aggregate, and they should have similar intentions and merge operations.

When an aggregate contains elements that refer to the same instance of another aggregate, it is necessary to define an intra-invariant to guarantee that they hold the same information. The tournament creator can also be a participant. Therefore, if the creator is added as participant while it is anonymized as creator, two concurrent versions of the aggregate are created having different information about the student. The merge would create an inconsistent tournament, but a correct intra-invariant, that states that creator and participant information should be the same, aborts the merge.

This experimentation allowed us to verify the use of the simulator on a large domain model, which addresses the second research question.

\subsection{Complexity Analysis}

In terms of the solution complexity, addressing the first research question, we compare with a microservices system implemented with eventual consistency and the Saga pattern~\cite{richardson19}. While the complexity of implementing the business logic using sagas depends on the number of distributed transactions and semantic locks~\cite{Santos20}, the complexity using transactional causal consistency depends on the number of aggregate elements, to define intentions and merge operations. The number of aggregates is usually smaller than the number of functionalities, and, in TCC, the propagation impact of adding a new aggregate is confined to the aggregate and its inter-invariants, while, in sagas, the impact of adding a new functionality may impact all other functionalities. However, more research and experimentation is required.

\subsection{Threats to Validity}

To simplify the explanation of transactional causal consistency, we used version numbers to select the aggregates that belong to a causal snapshot. However, in a distributed implementation the generation of a total order of version numbers is not possible, or at least it does not scale. Nevertheless, TCC~\cite{Wu20,lykhenko21} existing implementations support a similar semantics in the construction of causal snapshots using timestamp intervals. Additionally, the causal snapshot we defined is a best effort snapshot, which in the distributed context may be harder to achieve. Anyway, the transactional semantics and how it impacts on the business logic design is the same.

The simulator does not implement distributed communication in the invocations between the different services implementing a functionality. This simplifies the simulator and it does not impact the business logic, because each service invocation occurs in a different system transaction. Even though they are not distributed, the interleaving between service invocations can occur. On the other hand, it does not simulate distributed faults but this is not a concern of the simulation, because in TCC writes are atomic and if a failure occurs it only necessary retry the functionality. 

We are not addressing the replication of aggregates. This problem is usually understood as a low level data issue, associated with performance. However we look at it as a higher level of abstraction by defining the inter-invariants, which define consistency rules for the information replicated between two aggregates. This is, in our opinion, the correct way to look at it from a business logic perspective.

\section{Conclusions}
\label{sec:conclusions}

Microservices architectures have to handle the burden of eventual consistency. The Saga pattern has been used to implement the microservices functionalities, but it is well known that there is a trade-off between the amount of application business logic and its implementation effort using a microservices architecture. 

We leverage on previous work, that proposes the use of transactional causal consistency (TCC) in serverless computing, to define an approach for the use of TCC on the implementation of microservices business logic, which reduces its implementation complexity.

The approach proposes new aggregate constructs, which specify the intra and inter aggregate business logic, and their extension for transactional causal consistency. Additionally, due to the low level of existing TCC implementations, we designed and implemented a TCC simulator, which supports the new aggregate constructs. Finally, the constructs and their implementation using the simulator were experimented in a business logic rich application. Therefore, we have positively answered both research questions.

The simulator code and the case study implementation are publicly available at \url{https://github.com/socialsoftware/business-logic-consistency-models/tree/Pedro-Pereira-Thesis}.

\section*{Acknowledgment}
This work was partially supported by Fundação para a Ciência e Tecnologia (FCT) through projects UIDB/50021/2020 (INESC-ID) and PTDC/CCI-COM/2156/2021 (DACOMICO).

\bibliographystyle{IEEEtran}
\bibliography{bibliography}

\begin{thebibliography}{10}
\providecommand{\url}[1]{#1}
\csname url@samestyle\endcsname
\providecommand{\newblock}{\relax}
\providecommand{\bibinfo}[2]{#2}
\providecommand{\BIBentrySTDinterwordspacing}{\spaceskip=0pt\relax}
\providecommand{\BIBentryALTinterwordstretchfactor}{4}
\providecommand{\BIBentryALTinterwordspacing}{\spaceskip=\fontdimen2\font plus
\BIBentryALTinterwordstretchfactor\fontdimen3\font minus
  \fontdimen4\font\relax}
\providecommand{\BIBforeignlanguage}[2]{{%
\expandafter\ifx\csname l@#1\endcsname\relax
\typeout{** WARNING: IEEEtran.bst: No hyphenation pattern has been}%
\typeout{** loaded for the language `#1'. Using the pattern for}%
\typeout{** the default language instead.}%
\else
\language=\csname l@#1\endcsname
\fi
#2}}
\providecommand{\BIBdecl}{\relax}
\BIBdecl

\bibitem{Hanlon06}
C.~O'Hanlon, ``A conversation with werner vogels,'' \emph{Queue}, vol.~4,
  no.~4, p. 14–22, May 2006.

\bibitem{thones15}
J.~Th{\"o}nes, ``Microservices,'' \emph{IEEE Software}, vol.~32, no.~1, pp.
  116--116, 2015.

\bibitem{fowler_microservices}
M.~Fowler, ``Microservices,'' Web page:
  \url{http://martinfowler.com/articles/microservices.html}.

\bibitem{richardson19}
C.~Richardson, \emph{Microservices Patterns}.\hskip 1em plus 0.5em minus
  0.4em\relax Manning Publications Co., 2019.

\bibitem{Santos20}
N.~Santos and A.~Rito~Silva, ``A complexity metric for microservices
  architecture migration,'' in \emph{2020 IEEE International Conference on
  Software Architecture (ICSA)}, 2020, pp. 169--178.

\bibitem{Haywood17}
\BIBentryALTinterwordspacing
D.~Haywood, ``In defense of the monolith,'' \emph{Microservices vs. Monoliths -
  The Reality Beyond the Hype}, 2017. [Online]. Available:
  \url{https://www.infoq.com/minibooks/emag-microservices-monoliths/}
\BIBentrySTDinterwordspacing

\bibitem{Wu20}
\BIBentryALTinterwordspacing
C.~Wu, V.~Sreekanti, and J.~M. Hellerstein, ``Transactional causal consistency
  for serverless computing,'' in \emph{Proceedings of the 2020 ACM SIGMOD
  International Conference on Management of Data}, ser. SIGMOD '20.\hskip 1em
  plus 0.5em minus 0.4em\relax New York, NY, USA: Association for Computing
  Machinery, 2020, p. 83–97. [Online]. Available:
  \url{https://doi.org/10.1145/3318464.3389710}
\BIBentrySTDinterwordspacing

\bibitem{lykhenko21}
\BIBentryALTinterwordspacing
T.~Lykhenko, R.~Soares, and L.~Rodrigues, ``Faastcc: Efficient transactional
  causal consistency for serverless computing,'' in \emph{Proceedings of the
  22nd International Middleware Conference}, ser. Middleware '21.\hskip 1em
  plus 0.5em minus 0.4em\relax New York, NY, USA: Association for Computing
  Machinery, 2021, p. 159–171. [Online]. Available:
  \url{https://doi.org/10.1145/3464298.3493392}
\BIBentrySTDinterwordspacing

\bibitem{evans03}
E.~Evans, \emph{Domain-Driven Design: Tackling Complexity in the Heart of
  Software}.\hskip 1em plus 0.5em minus 0.4em\relax Addison Wesley, 2003.

\bibitem{Fox99}
A.~Fox and E.~A. Brewer, ``Harvest, yield, and scalable tolerant systems,'' in
  \emph{Proceedings of the The Seventh Workshop on Hot Topics in Operating
  Systems}, ser. HOTOS '99.\hskip 1em plus 0.5em minus 0.4em\relax USA: IEEE
  Computer Society, 1999, p. 174.

\bibitem{bailis13}
P.~Bailis and A.~Ghodsi, ``Eventual consistency today: Limitations, extensions,
  and beyond,'' \emph{Communications of the ACM}, vol.~56, no.~5, pp. 55--63,
  2013.

\bibitem{Garcia-Molina87}
H.~Garcia-Molina and K.~Salem, ``Sagas,'' in \emph{Proceedings of the 1987 ACM
  SIGMOD International Conference on Management of Data}, ser. SIGMOD
  '87.\hskip 1em plus 0.5em minus 0.4em\relax New York, NY, USA: Association
  for Computing Machinery, 1987, p. 249–259.

\bibitem{Mendonca21}
N.~C. Mendonça, C.~Box, C.~Manolache, and L.~Ryan, ``The monolith strikes
  back: Why istio migrated from microservices to a monolithic architecture,''
  \emph{IEEE Software}, vol.~38, no.~5, pp. 17--22, 2021.

\bibitem{akkoorath16}
D.~D. Akkoorath, A.~Z. Tomsic, M.~Bravo, Z.~Li, T.~Crain, A.~Bieniusa,
  N.~Preguiça, and M.~Shapiro, ``Cure: Strong semantics meets high
  availability and low latency,'' in \emph{2016 IEEE 36th International
  Conference on Distributed Computing Systems (ICDCS)}, 2016, pp. 405--414.

\bibitem{braun21}
\BIBentryALTinterwordspacing
S.~Braun, A.~Bieniusa, and F.~Elberzhager, ``Advanced domain-driven design for
  consistency in distributed data-intensive systems,'' in \emph{Proceedings of
  the 8th Workshop on Principles and Practice of Consistency for Distributed
  Data}, ser. PaPoC '21.\hskip 1em plus 0.5em minus 0.4em\relax New York, NY,
  USA: Association for Computing Machinery, 2021. [Online]. Available:
  \url{https://doi.org/10.1145/3447865.3457969}
\BIBentrySTDinterwordspacing

\bibitem{Preguica2009}
N.~Pregui{\c c}a, J.~M. Marques, M.~Shapiro, and M.~Letia, ``A commutative
  replicated data type for cooperative editing,'' in \emph{29th IEEE
  International Conference on Distributed Computing Systems}, ser. ICDCS
  2009.\hskip 1em plus 0.5em minus 0.4em\relax IEEE, 2009, pp. 395--403.

\bibitem{yu20}
\BIBentryALTinterwordspacing
W.~Yu and C.-L. Ignat, ``{Conflict-Free Replicated Relations for
  Multi-Synchronous Database Management at Edge},'' in \emph{{IEEE
  International Conference on Smart Data Services, 2020 IEEE World Congress on
  Services}}, Beijing, China, Oct. 2020. [Online]. Available:
  \url{https://hal.inria.fr/hal-02983557}
\BIBentrySTDinterwordspacing

\bibitem{fowler03}
M.~Fowler, \emph{Patterns of Enterprise Application Architecture}.\hskip 1em
  plus 0.5em minus 0.4em\relax Addison-Wesley, 2003.

\bibitem{clements11}
P.~Clements, F.~Bachmann, L.~Bass, D.~Garlan, J.~Ivers, R.~Little, P.~Merson,
  R.~Nord, and J.~Stafford, \emph{Documenting Software Architectures: Views and
  Beyond 2nd Edition}.\hskip 1em plus 0.5em minus 0.4em\relax Addison-Wesley,
  2011.

\end{thebibliography}

\end{document}